\documentclass[showpacs,amsmath,amssymb,aps,showkeys,floatfix,prd,a4paper]{revtex4}

\usepackage{graphicx}
\usepackage{dcolumn}
\usepackage{bm}
\usepackage{amsfonts}
\usepackage{amssymb,amscd}

\def\pythia{\textsc{Pythia}}
\def\lsim{\raise0.3ex\hbox{$<$\kern-0.75em\raise-1.1ex\hbox{$\sim$}}}
\def\gsim{\raise0.3ex\hbox{$>$\kern-0.75em\raise-1.1ex\hbox{$\sim$}}}

\def\pom{{I\!\!P}}

\def\beq{\begin{equation}}
\def\eeq{\end{equation}}
\def\bea{\begin{eqnarray}}
\def\eea{\end{eqnarray}}
\def\bq{\begin{quote}}
\def\eq{\end{quote}}

\def\gappeq{\mathrel{\rlap {\raise.5ex\hbox{$>$}}
{\lower.5ex\hbox{$\sim$}}}}

\def\lappeq{\mathrel{\rlap{\raise.5ex\hbox{$<$}}
{\lower.5ex\hbox{$\sim$}}}}

\def\Toprel#1\over#2{\mathrel{\mathop{#2}\limits^{#1}}}

\def\pom{{I\!\!P}}

\begin{document}

%\preprint{Version 1.2}

\title{Diffractive gauge boson production at the LHC as a probe of the flavour content of the Pomeron}

\author{E. A. F. Basso$^{1}$, V. P. Goncalves$^{2}$, D. E. Martins$^3$,  M. S. Rangel$^{3}$}

\affiliation{$^{1}$ $^{1}$ Faculdade de Ci\^encias Exatas e Tecnologia,
Universidade Federal da Grande Dourados (UFGD),\\
Caixa Postal 364, Dourados, CEP 79804-970, MS, Brazil.}

\affiliation{$^{2}$ Instituto de F\'{\i}sica e Matem\'atica,  Universidade
Federal de Pelotas (UFPel), \\
Caixa Postal 354, CEP 96010-900, Pelotas, RS, Brazil}

\affiliation{$^{3}$ Instituto de F\'isica, Universidade Federal do Rio de Janeiro (UFRJ), 
Caixa Postal 68528, CEP 21941-972, Rio de Janeiro, RJ, Brazil}

\date{\today}

\begin{abstract}
The production of the $W$ and $Z$ bosons in single diffractive processes at the LHC is investigated taking into account the ATLAS, CMS and LHCb acceptances and considering different assumptions for the flavour content of the Pomeron. The total cross sections and pseudorapidity distributions are estimated for $pp$ collisions at $\sqrt{s} = 13$ TeV. Our results indicate that  a future experimental analysis of the ratio between the $W$ and $Z$ cross sections can be used to probe the flavour content of the Pomeron.

\end{abstract}

%\pacs{12.38.-t; 13.60.Le; 13.60.Hb}

\maketitle

The study of hadronic collisions at the LHC provides a unique environment for precise measurements of poorly understood phenomena. In particular, the study of hard diffractive processes at the LHC is expected to provide important insight for improving the theoretical description of the diffractive physics and the nature of the {Pomeron} ($\mathbb {P}$),  which is a long-standing puzzle in  Particle Physics \cite{pomeron}. This object, with the vacuum quantum numbers, was introduced phenomenologically in the Regge theory as a simple moving pole in the complex angular momentum plane, to describe the high-energy behaviour of the total and elastic cross-sections of the hadronic reactions. On the other hand, the diffractive 
deep inelastic scattering (DDIS) at HERA is quite well described assuming the validity Regge factorization of the diffractive processes, 
as suggested long ago by Ingelman and Schlein (IS) \cite{Ingelman:1984ns} and 
not yet proven in pQCD. The IS approach essentially considers 
that the diffractive processes can be described in terms of the probability of 
the proton to emit a colour singlet object -- the  Pomeron -- and the subsequent 
interaction of the partons inside such Pomeron with the virtual photon emitted 
by the incident electron. This introduces the Pomeron parton distribution 
functions, which can be extracted from data where a hard final state is 
produced and a leading hadron is detected. During the last years, several groups have used the HERA data to extract the gluon and quark distributions of the Pomeron considering different assumptions for the initial conditions and the DGLAP evolution at leading and next - to - leading orders 
\cite{martin_dif,royon_dif, H1diff,H1Diff2,zeus_dif,ira_dif,guzey_dif}. The main conclusion of these different analyzes is that the Pomeron structure is dominated by gluons, with the quark content being non - negligible. One important assumption present in these studies is that the flavour content of the Pomeron is equal for up, down and strange quarks, i.e. $u_{\pom} = d_{\pom} = s_{\pom} = q_{\pom}$ with $q_{\pom} = \bar{q}_{\pom}$. Such assumption arise due to the fact
 that the HERA data do not allow to separate the contribution of the distinct light quarks for the Pomeron structure. Our goal in this paper is to demonstrate that the diffractive gauge boson production at the LHC can improve our understanding of the
 flavour content of the Pomeron.

\begin{figure}[!tb]
\includegraphics[width=0.45\textwidth]{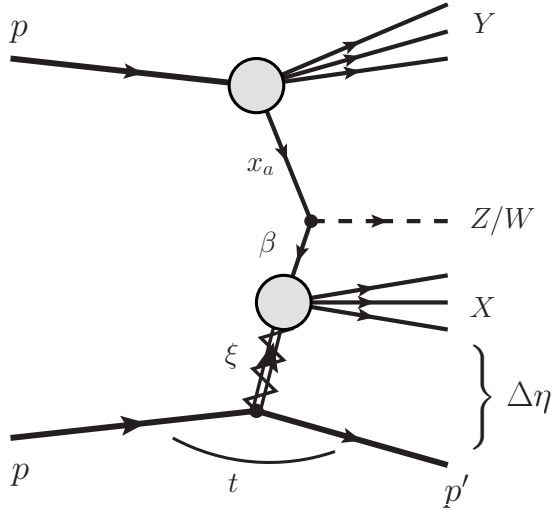}
\caption{Single diffractive production of gauge bosons in $pp$ collisions.}
\label{fig:Diagrama}
\end{figure}

The recent studies of $W^{\pm}$ and $Z^0$ production in hadronic collisions  are  in general dedicated to the calculation of the production of this final state in inclusive reactions, where both initial protons  dissociate in the interaction.   However, these gauge bosons   can also be produced in diffractive interactions, where one (or both) of the protons remain intact and   empty regions  in pseudo-rapidity, called rapidity gaps, separate the intact very forward proton(s) from the gauge boson state (For previous studies see, e.g. Refs. 
  \cite{Cov,Gay,Gol,roy,Ing,royon,torbcris,nos_gauge}). In this paper we will focus in the gauge boson production in single diffractive processes,  represented in  Fig. \ref{fig:Diagrama}. In the IS approach, denoted often as Resolved Pomeron Model, the Pomeron is assumed to have a 
well defined partonic structure and the hard process takes place in a 
Pomeron - proton or proton - Pomeron interaction in the case of single 
diffractive processes. At leading order the gauge boson production  is 
determined by the annihilation processes $q \bar q \rightarrow G$ ($G = W$ 
or $Z$). Higher order contributions are not considered here and can be taken 
into account effectively by a $K$ factor. In order to estimate the hadronic 
cross sections we have to convolute the cross section for this partonic 
subprocess with the inclusive and diffractive parton distribution functions. 
In the collinear factorization formalism, the single diffractive gauge boson 
production cross section can be expressed by
\begin{eqnarray}
\sigma^{SD}(pp \rightarrow Y GX \otimes p) = \sum_{a,b} \int dx_a \int dx_b 
\left[f^D_a(x_a,\mu^2) f_b(x_b,\mu^2) + f_a(x_a,\mu^2)f^D_b(x_b,\mu^2)\right] 
\, \hat{\sigma}_{ab \rightarrow G} \,\,, \label{cross} 
\end{eqnarray}
where $\otimes$ represents a rapidity gap in the final state, $Y$ the product of the proton dissociation, $X$ the Pomeron remnants and $f_i(x_i,\mu^2)$ and $f_i^D(x_i,\mu^2)$ are the inclusive and diffractive 
parton distribution functions, respectively. In our study  we use the inclusive parton distributions as given by the 
CT10 parametrization \cite{ct10}. In Eq. (\ref{cross}) the $p \pom$ and 
$\pom p$ interactions are included. Moreover, $\hat{\sigma}_{ab\rightarrow G}$ denotes the 
hard partonic interaction producing a gauge boson. The Resolved Pomeron Model considers the diffractive quark  distributions as a 
convolution of the Pomeron flux emitted by the proton, $f_{\pom}(x_{\pom})$, 
and the parton distributions in the Pomeron,  
$q_{\pom}(\beta, \mu^2)$, where $\beta$ is the momentum fraction carried by the 
partons inside the Pomeron. The Pomeron flux is given by 
\begin{eqnarray}
f^{p}_{\pom}(\xi)= 
\int_{t_{\rm min}}^{t_{\rm max}} dt \, \frac{A_{\pom} \, e^{B_{\pom} t}}{\xi^{2\alpha_{\pom} (t)-1}} \,\,.
\label{fluxpom:proton}
\end{eqnarray}
The normalization of the flux is such that the relation 
$\xi \int^{t_{\text{min}}}_{{t_\text{cut}}} \, dt f_{\pom/p}(\xi,t) = 1$ 
holds at $\xi = 0.003$, where $|t_\text{cut}| = 1$ GeV is limited by the 
measurement and $|t_\text{min}| \simeq (m_\text{p} \xi)^2/(1-\xi)$ is the 
kinematic limit of accessible momentum $|t|$. The Pomeron flux factor is 
motivated by Regge theory, where the Pomeron trajectory assumed to be linear, 
$\alpha_{\pom} (t)= \alpha_{\pom} (0) + \alpha_{\pom}^\prime t$, and the 
parameters $B_{\pom}$, $\alpha_{\pom}^\prime$ and their uncertainties are 
obtained from fits to H1 data  \cite{H1diff}.
In the present analysis the H1 Fit B is used, 
which has the slope parameter set to $B_\pom = 5.5$ GeV${}^{-2}$ and 
$\alpha_\pom^\prime = 0.06$ GeV${}^{-2}$, while $\alpha_\pom(0) = 1.111 \pm 
0.007 $. Consequently,  the diffractive quark distributions are  given by
\begin{eqnarray}
{ q^D(x,\mu^2)}=\int d\xi d\beta \delta (x-\xi \beta)f_{\pom}(\xi)
q_{\pom}(\beta, \mu^2)={ \int_x^1 \frac{d\xi}{\xi} f_{\pom}(\xi) 
q_{\pom}\left(\frac{x}{\xi}, \mu^2\right)}\,.
\end{eqnarray}
The quark distributions of the Pomeron have been extracted from the HERA DIS measurements for the diffractive proton structure function $F_2^{D(4)}$ assuming that
$u_{\pom} = d_{\pom} = s_{\pom} = q_{\pom}$, with $q_{\pom} = \bar{q}_{\pom}$. However, at leading order we have that
\begin{eqnarray}
\rm{F}_{2}^{\rm{D}(4)} \propto \left(\frac{2}{3}\right)^{2}u_{\mathbb{P}} + \left(\frac{1}{3}\right)^{2}d_{\mathbb{P}}+ \left(-\frac{1}{3}\right)^{2}s_{\mathbb{P}}\,\,
\label{F2D}
\end{eqnarray}
which implies that the constrain $4u_{\mathbb{P}} + d_{\mathbb{P}}+ \bar{s}_{\mathbb{P}}=6q_{\pom}$ must be satisfied. Defining the auxiliary functions
\begin{eqnarray}
R_{ud}= \frac{u_{\mathbb{P}}}{d_{\mathbb{P}}}, \,\,\,\, R_{sd}=\frac{s_{\mathbb{P}}}{d_{\mathbb{P}}}\,\,, 
\end{eqnarray}
the diffractive PDFs can be expressed as follows: 
\begin{eqnarray}
u_{\mathbb{P}}(\beta,\mu^{2})&=&  \frac{6R_{ud}}{1+ R_{sd}+ 4R_{ud}}\cdot q_{\pom}(\beta,\mu^{2}) \nonumber \\ 
d_{\mathbb{P}}(\beta,\mu^{2})&=&  \frac{6}{1+ R_{sd}+ 4R_{ud}}\cdot q_{\pom}(\beta,\mu^{2}) \nonumber \\
s_{\mathbb{P}}(\beta,\mu^{2})&=&  \frac{6R_{sd}}{1+ R_{sd} + 4R_{ud}}\cdot q_{\pom}(\beta,\mu^{2}).  
\end{eqnarray}
For  $R_{ud}=R_{sd}=1$ we recover the default HERA diffractive distributions. In order to test the dependence of the gauge boson production on the flavour content of the Pomeron, in what follows we  will consider some different assumptions for the value of the ratios $R_{ud}$ and $R_{sd}$, which we  assume to be scale independent. In particular, as in Ref. \cite{royon}, we will consider that these ratios can also assume, independently, the values 0.5 and 2.0, and will compare with the default predictions. As demonstrated in Ref. \cite{royon}, these different assumptions has direct impact on the $W$ charge asymmetry. Our goal is to extend this previous analysis for the $Z$ boson production and present, for the first time, predictions for the rapidity distributions and ratio between cross sections considering the kinematic range probed by the ATLAS, CMS and LHCb detectors and the typical cutoffs present in the selection of the single diffractive events.

Before to present our results, some comments are in order. First, at large 
values of the Pomeron momentum fraction $\xi$, subleading contributions 
associated to Reggeon exchange can be important in some regions of the phase 
space (For a discussion see e.g. Ref. \cite{marquet,royon}). 
In what follows we will focus our analysis in the kinematical region where $\xi \le 0.12$, in which the Reggeon contribution is negligible. Second, in order to estimate the diffractive cross sections in $pp$ collisions, one also have to take into account of the soft rescattering corrections associated to reinteractions (often referred to as multiple scatterings) between spectator partons of the colliding proton that implies   the breakdown of the factorization assumed in Eq. (\ref{cross}) \cite{bjorken}. 
 One 
possible approach to treat this problem is to assume that the hard process 
occurs on a short enough timescale such that the physics that generate the 
additional particles can be factorized and accounted by an overall factor, 
denoted gap survival factor $\langle |S|^2\rangle$, multiplying the cross section 
calculated using the collinear factorization and the diffractive parton 
distributions extracted from HERA data. The modelling and magnitude of this 
factor still is a theme of intense debate \cite{durham,telaviv}. In 
general the values of $\langle |S|^2\rangle$   depend on the energy, being typically of 
order $1-5$\% for LHC energies. Such 
approach have been largely used in the literature to estimate the hard 
diffractive processes at the LHC with reasonable success 
to describe the current data. In what follows we will assume the validity of this approach, with $\langle |S|^2\rangle = 0.05$ for single diffractive processes (For a more detailed discussion see e.g. \cite{nos_gauge}).

The single diffractive gauge boson production at the LHC will be estimated using  the Forward Physics Monte Carlo (\texttt{FPMC}) event generator~\cite{fpmc}. This Monte Carlo allows  to produce event samples for the diffractive W $\rightarrow \nu \mu$ and Z $\rightarrow \mu \mu$ processes and to obtain realistic predictions for the boson W$^{\pm}$ and Z production with one leading intact hadron, taking into account the acceptance of the LHC detectors. The distributions are obtained for $pp$ collisions at $\sqrt{s}=13\,\mbox{TeV}$ considering the  LHCb, CMS and ATLAS acceptances and a non diffractive background in the W$^{\pm}$ and Z production. The events have been generated assuming the following combinations for the ratios $R_{ud}=u_{\mathbb{P}}/d_{\mathbb{P}}$ and $R_{sd}=s_{\mathbb{P}}/d_{\mathbb{P}}$: (0.5,0.5), (0.5,1), (2,1) and (2,2). An integrated luminosity of 100 $\rm{pb}^{-1}$ (CMS and ATLAS) and 6 $\rm{fb}^{-1}$ (LHCb) is assumed. 
In the case of the CMS and ATLAS detectors, the following selection criteria have been considered: the muons must have $p_{T}\left(\mu^{\pm}\right)> 30\mbox{GeV}$ at the central region inside the interval $\left|\eta\left(\mu^{\pm}\right)\right|<2.4$ and  the transverse mass of the $W$ bosons, given by, $M_{T} = \sqrt{(E_{T,\mu}+ E_{T,\nu_{\mu}})^{2} - (p_{T,\mu}+ p_{T,\nu_{\mu}})^{2}}$, is required to be greater than 40 GeV.
% {\bf tagging of the proton in the final state??}.
  On the other hand, in the case of the LHCb detector, we assume that 
the muons with $p_{T}\left(\mu^{\pm}\right)> 20\mbox{GeV}$ must be at the forward region inside the pseudorapidity window of $2.0<\eta\left(\mu^{\pm}\right)<4.5$. Moreover, a VELO gap requirement in the backward region is performed using charged particles with momentum greater than 100 MeV inside the rapidity range  $-1.5 > \eta > -3.5$ acceptance. Finally, a HERSCHEL gap requirement in the backward region is performed using charged and neutral particles with momentum greater than 500 MeV inside the $-5.5 > \eta > -8.0$  acceptance~\cite{herschel}.

\begin{center}
\begin{table}[!t]
\begin{tabular}{ l||c|c|c||c|c|c| }
\hline 
\hline
  & \multicolumn{3}{c}{{ $\sigma^{SD}(pp \rightarrow W^{+} p)$ [pb]}}  & \multicolumn{3}{c}{{$\sigma^{SD} (p p \rightarrow Z p)$ [pb]}} \\
\hline
Flavour content & Full rapidity range & ATLAS/CMS & LHCb & Full rapidity range & ATLAS/CMS & LHCb \\
 \hline
 $u_{\mathbb{P}}=s_{\mathbb{P}}= d_{\mathbb{P}}$ & 115.8 & 13.6   & 2.0  & 59.7  & 3.1  & 0.18  \\
 \hline 
 $u_{\mathbb{P}}/d_{\mathbb{P}}= 0.5, s_{\mathbb{P}}/d_{\mathbb{P}}=0.5$ & 147.1  & 18.0  & 3.2  & 61.3  & 3.7  & 0.21  \\
 \hline 
 $u_{\mathbb{P}}/d_{\mathbb{P}}= 0.5, s_{\mathbb{P}}/d_{\mathbb{P}}=1$ & 142.2 & 17.6  & 2.9  & 58.6  & 3.9 & 0.20  \\
 \hline 
 $u_{\mathbb{P}}/d_{\mathbb{P}}= 2, s_{\mathbb{P}}/d_{\mathbb{P}}=1$ & 94.9 & 9.1  & 1.3  & 60.5  & 2.5  & 0.17 \\
 \hline 
 $u_{\mathbb{P}}/d_{\mathbb{P}}= 2, s_{\mathbb{P}}/d_{\mathbb{P}}=2$ & 96.0  & 9.0  & 1.4  & 58.5  & 2.7  & 0.16 \\
 \hline 
\hline  
 \end{tabular}
\caption{Predictions for the single diffractive cross sections for the $W^+$ and $Z$ production considering the ATLAS, CMS and LHCb acceptances and different assumptions for the flavour content of the Pomeron. }
\label{table:sigma}
\end{table}
\end{center}

\begin{figure}[t]
\includegraphics[width=0.475\textwidth]{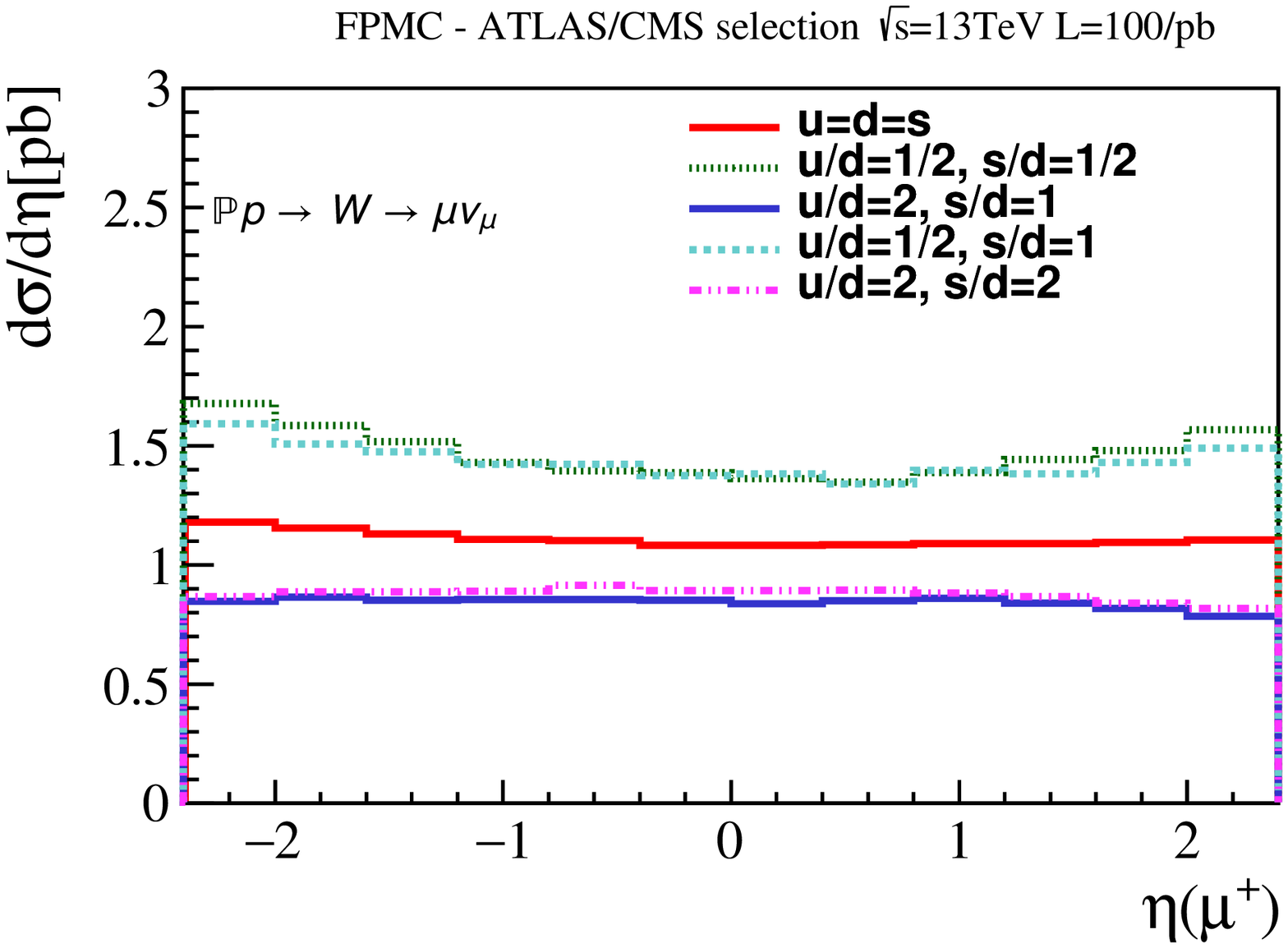}
\includegraphics[width=0.475\textwidth]{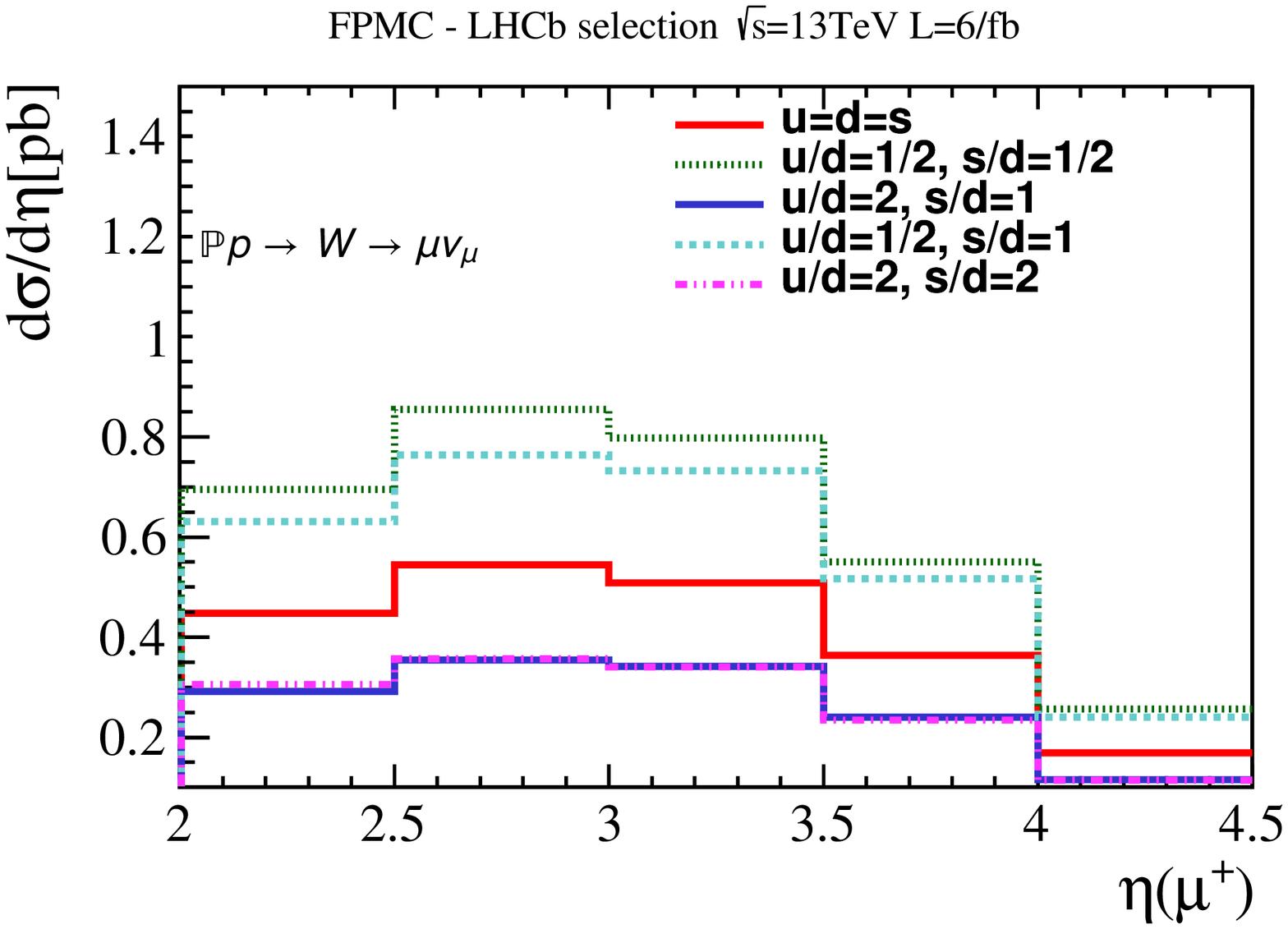}
\caption{Differential cross-section as function of $\eta(\mu^{+})$ for the single diffractive $W^+$ production 
in $pp$ collisions at $\sqrt{s}=13$ TeV considering the ATLAS/CMS (left) and LHCb (right) acceptances and different assumptions for the flavour content of the Pomeron.}
\label{fig:1}
\end{figure}

\begin{figure}[t]
\includegraphics[width=0.475\textwidth]{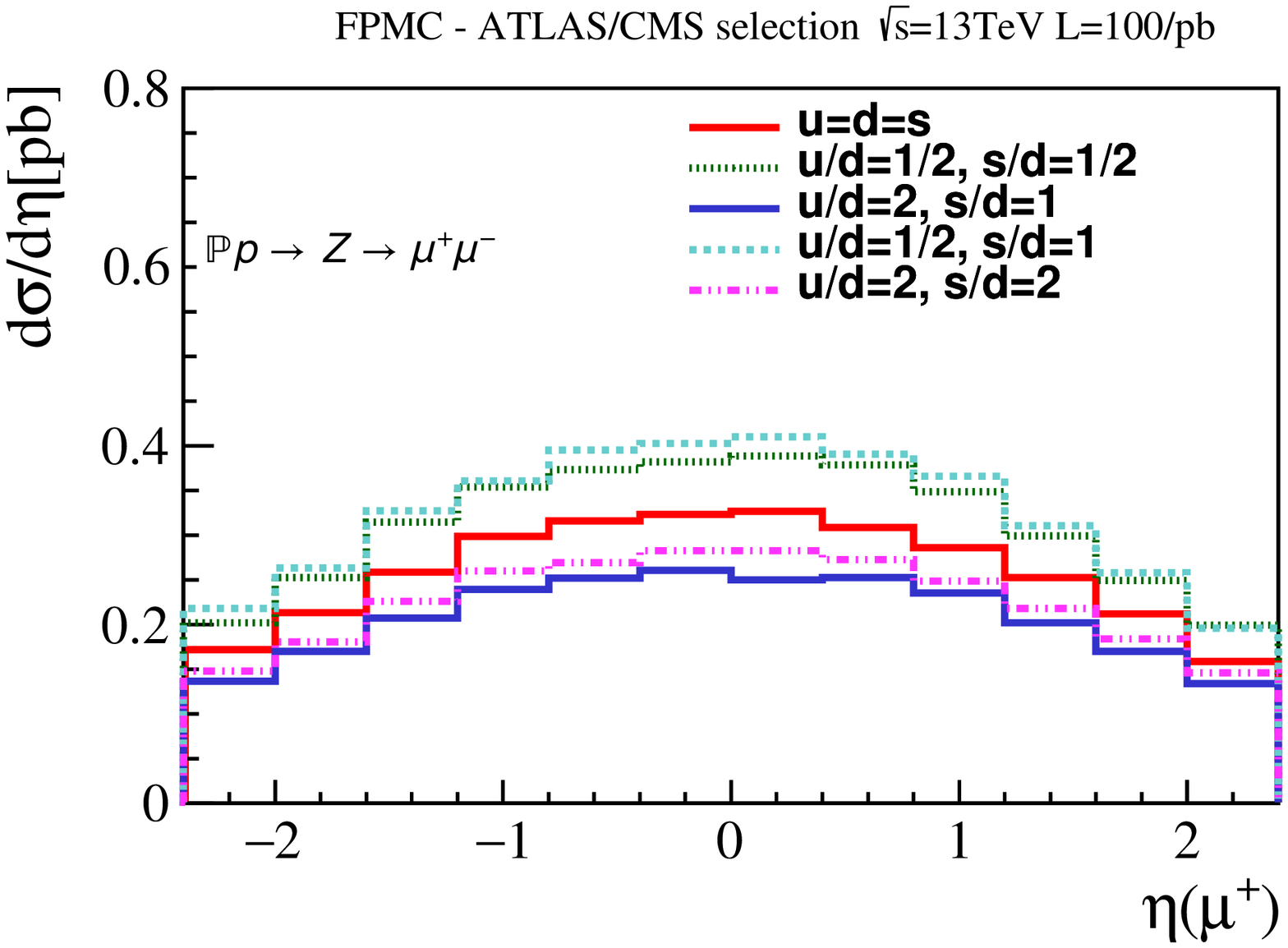}
\includegraphics[width=0.475\textwidth]{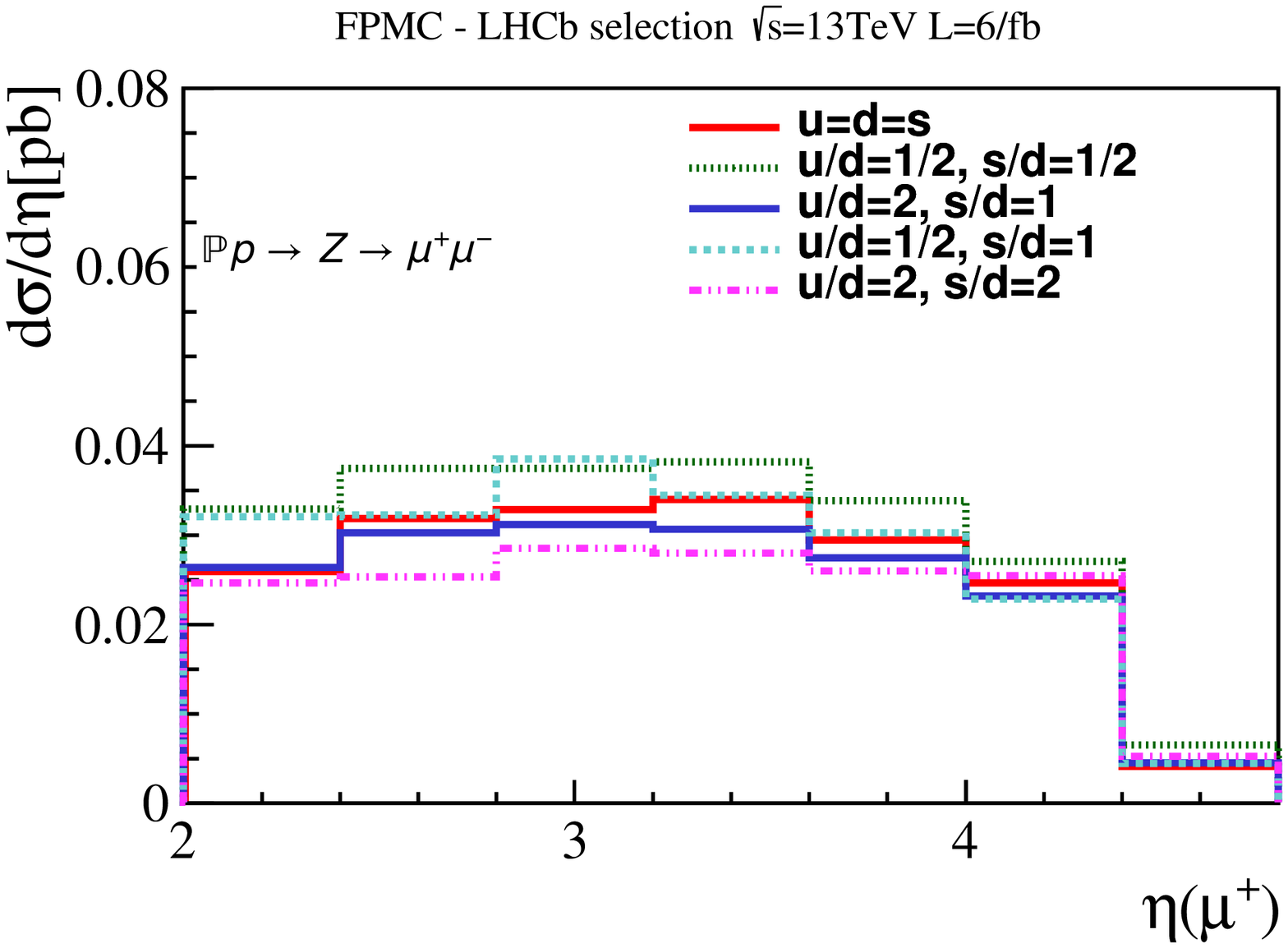}
\caption{Differential cross-section as function of $\eta(\mu^{+})$ for the single diffractive $Z$ production 
in $pp$ collisions at $\sqrt{s}=13$ TeV considering the ATLAS/CMS (left) and LHCb (right) acceptances and different assumptions for the flavour content of the Pomeron.}
\label{fig:2}
\end{figure}

In Table \ref{table:sigma} we present our predictions for the total cross sections considering the acceptances of the ATLAS, CMS and LHCb detectors and different assumptions for the flavour content of the Pomeron. For comparison we also present the results for the full LHC rapidity range.
% {?? \bf without the inclusion of the selection criteria ??}
We have that the predictions for the $W^+$  production can differ by a factor $\lesssim 3$ depending of the values for $R_{ud}$ and $R_{sd}$. For $Z$ production, the difference between the predictions is smaller than a factor 1.5 . Moreover, our results indicate that the cross sections are not strongly sensitive to the ratio $R_{sd}$. Such conclusion is also obtained from the analysis of  the results presented in Fig. 
\ref{fig:1} for the pseudorapidity $\mu^+$ distributions. We have that the reduction of $u$ quarks in the Pomeron, and corresponding enhancement of $d$ quarks, present in the assumption $R_{ud} = 1/2$ imply a larger cross section in comparison to the default $R_{ud} = 1$ one. On the other hand, $R_{ud} = 2$ imply a smaller cross section.
Similar behaviour also is present in the single diffractive $Z$ production, as observed in the results presented in Fig. \ref{fig:2}. However, the impact of the different assumptions for the flavour content of the Pomeron is smaller, in agreement with the results presented in the Table \ref{table:sigma}. We have that the shape of the distributions is not sensitive to these assumptions. That is an important shortcoming to probe the flavour content of the Pomeron, due to the current theoretical uncertainty on the value of the absorptive corrections for the diffractive processes. 

\begin{figure}[t]
\includegraphics[width=0.475\textwidth]{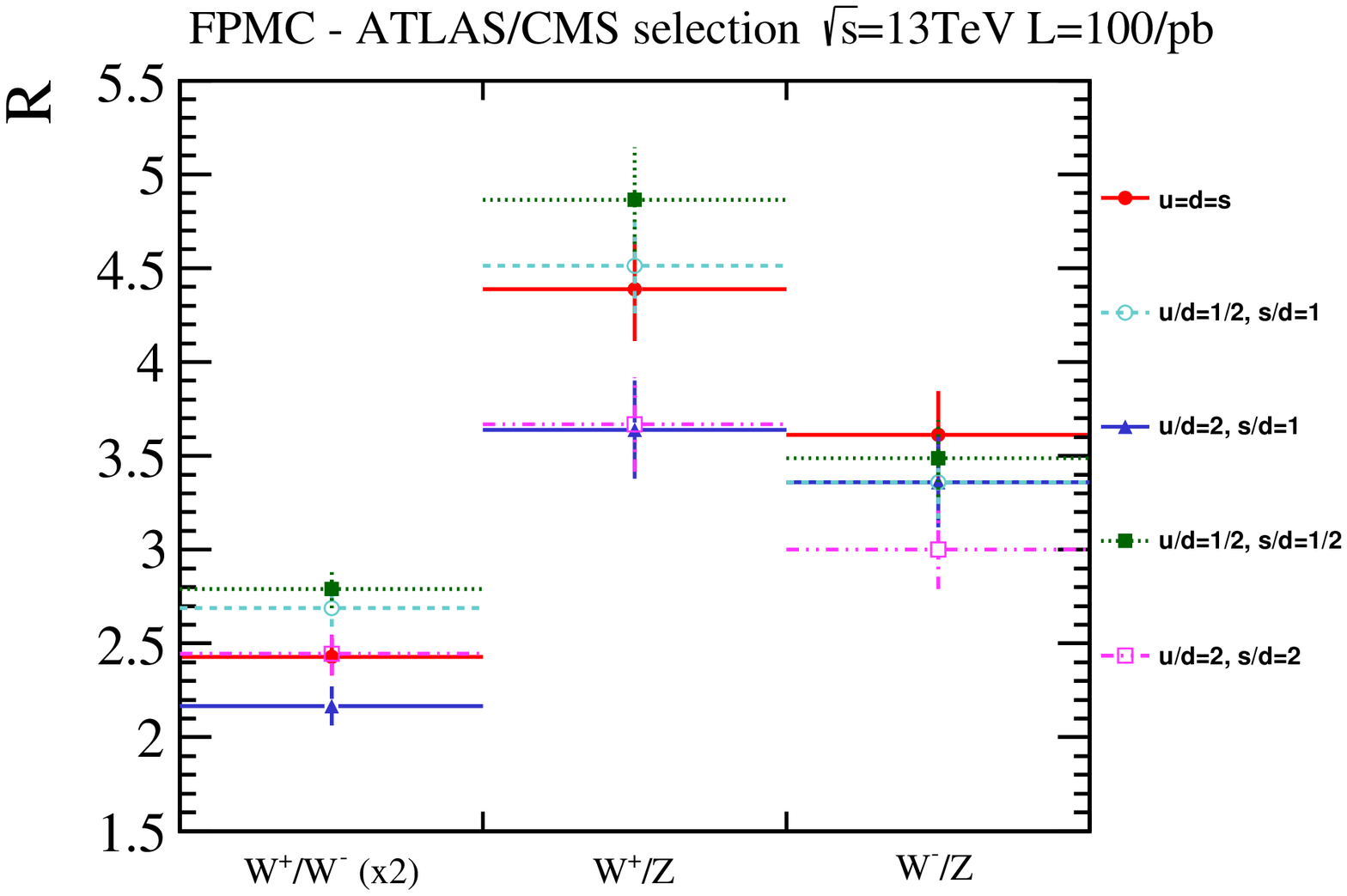}
\includegraphics[width=0.475\textwidth]{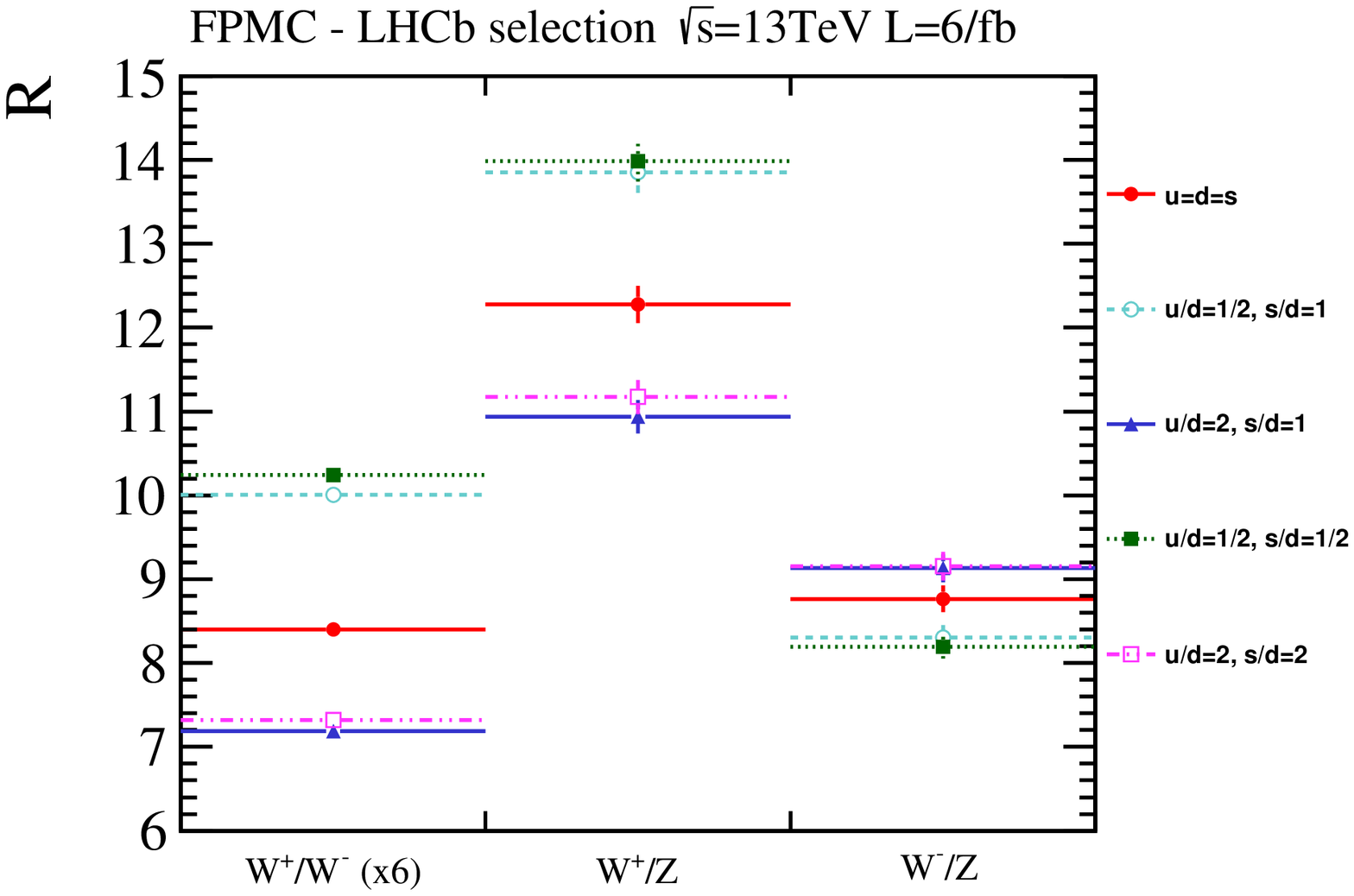}
\caption{Predictions for the ratios 
${\sigma^{SD}_{W^{+}}}/{\sigma^{SD}_{W^{-}}}$,  ${\sigma^{SD}_{W^{+}}}/{\sigma^{SD}_{Z}}$ and ${\sigma^{SD}_{W^{-}}}/{\sigma^{SD}_{Z}}$ considering the ATLAS/CMS (left) and LHCb (right) acceptances. The CMS/ATLAS (LHCb) predictions for ${\sigma^{SD}_{W^{+}}}/{\sigma^{SD}_{W^{-}}}$ have been rescaled by a factor  2 (6) to allow the comparison with the other ratios.}
\label{fig:3}
\end{figure}

An alternative to surpass this shortcoming is to consider the ratio between cross sections, which cancel several of the experimental and theoretical systematic uncertainties. In particular, as the absorptive corrections are expected to be similar for the diffractive $W^{\pm}$ and $Z$ production, such ratios should not be sensitive to the modelling of these effects.
In Fig. \ref{fig:3} 
we present our predictions for the ratios ${\sigma^{SD}_{W^{+}}}/{\sigma^{SD}_{W^{-}}}$  ${\sigma^{SD}_{W^{+}}}/{\sigma^{SD}_{Z}}$ and ${\sigma^{SD}_{W^{-}}}/{\sigma^{SD}_{Z}}$ considering the ATLAS/CMS (left) and LHCb (right) acceptances. 
Since LHCb has no forward proton detectors, we included in the ratio predictions the non-diffractive contribution as predicted by \pythia~8~\cite{rpythia}. Our results indicate that the ratios between the $W$ and $Z$ cross sections are sensitive to the flavour content of the Pomeron, with the magnitude being dependent of the charge of the $W$ boson.
From the analysis of these results, we propose the measurement of both ratios, 
${\sigma^{SD}_{W^{+}}}/{\sigma^{SD}_{Z}}$ and ${\sigma^{SD}_{W^{-}}}/{\sigma^{SD}_{Z}}$, as a strategy the constrain the modelling of the flavour content of the Pomeron. 

Finally, let's summarize our main results and conclusions. The description of the diffractive processes is still an important open question. In particular, the quark content of the Pomeron has been poorly constrained by the HERA data. In this paper we have investigated the single diffractive gauge boson production as a probe of the flavour content of the Pomeron. We have used the Forward Physics Monte Carlo and obtained realistic predictions for the single diffractive $W^+$ and $Z$ production taking into account the acceptance of the ATLAS, CMS and LHCb detectors. In the case of the ATLAS and CMS detectors we have assumed the tagging of one of the protons in the final state, which allow the direct separation of the single diffractive events. On the other hand, in the LHCb case, we have considered a gap requirement in the VELO and HERSCHEL detectors. Our results indicate that the magnitude of the distributions is sensitive to the assumptions about the content of $u$, $d$ and $s$ quarks in the Pomeron. As the shape of the distributions are not strongly modified by these assumptions, we have proposed the analysis of the ratio between cross sections in order to reduce the impact of the absorptive corrections in our predictions. Our results indicate that a future experimental analysis of the ratio between the $W$ and $Z$ cross sections can be used to probe the flavour content of the Pomeron.

\section*{Acknowledgements}
The authors acknowledge useful discussions with C.~Royon in the initial phase of this project.
 VPG acknowledges useful discussions with  J.~D.~Tapia~Takaki and is grateful to the members of the Department of Physics and Astronomy of the University of Kansas by the warm hospitality during the development of part of this study.
This work was  partially financed by the Brazilian funding agencies CNPq, CAPES, FAPERJ, FAPERGS and INCT-FNA (process number 464898/2014-5).

%%%%%%%%%%%%%%%%%%%%%%%%%%%%%%%%%%%%%%%%%%%%%%%%%%%%%%%%%%%%%%%%%%%%%%%%%%%%%%%%

%%%%%%%%%%%%%%%%%%%%%%%%%%%%%%%%%%%%%%%%%%%%%%%%%%%%%%%%%%%%%%%%%%%%%%%%%%%%%%%%

\end{document}